\documentstyle[aaspp4,12pt,amsmath]{article}

\lefthead{FRYER, HOLZ, \& HUGHES} 

\righthead{Gravitational Wave Emission From Collapse}

\received{2004 January 6}
\begin{document} 




\title{Gravitational Waves from Stellar Collapse:  Correlations 
to Explosion Asymmetries} 
\author{Chris L. Fryer} 
\affil{Theoretical Astrophysics, 
Los Alamos National Laboratories, \\ Los Alamos, NM 87544}
\authoremail{clf@t6-serv.lanl.gov}

\author{Daniel E. Holz}
\affil{Center for Cosmological Physics \\
University of Chicago \\
Chicago, IL  60637}
\authoremail{deholz@itp.ucsb.edu}

\author{Scott A. Hughes}
\affil{Department of Physics \\
Massachusetts Institute of Technology \\
Cambridge, MA 02139-4307}
\authoremail{hughes@itp.ucsb.edu}

\begin{abstract}
  
  The collapse of massive stars not only produces observable outbursts
  across the entire electromagnetic spectrum but, for Galactic (or
  near-Galactic) supernovae, detectable signals for ground-based
  neutrino and gravitational wave detectors.  Gravitational waves and
  neutrinos provide the only means to study the actual engine behind
  the optical outbursts: the collapsed stellar core.  While the
  neutrinos are most sensitive to details of the equation of state,
  gravitational waves provide a means to study the mass asymmetries in
  this central core.  We present gravitational wave signals from a
  series of 3-dimensional core-collapse simulations with asymmetries
  derived from initial perturbations caused by pre-collapse
  convection, core rotation, and low-mode convection in the explosion
  engine itself.  A Galactic supernovae will allow us to differentiate
  these different sources of asymmetry.  Combining this signal with
  other observations of the supernova, from neutrinos to gamma-rays to
  the compact remnant, dramatically increases the predictive power of
  the gravitational wave signal.  We conclude with a discussion of the
  gravitational wave signal arising from collapsars, the leading
  engine for long-duration gamma-ray bursts.

\end{abstract}


\section{Introduction}

The mechanism behind core-collapse supernovae (SNe) remains one of the
longest outstanding problems in astrophysics.  Whether or not
simulations of stellar collapse produce explosions depends sensitively
on the numerical implementations of the neutrino transport and the
equation of state physics (Herant et al. 1994; Burrows, Hayes, Fryxell
1995; Janka \& M\"uller 1996; Mezzacappa et al. 1998; Fryer \& Warren
2002; Buras et al. 2003).  10\% differences in this physics can make
the difference between the success or failure of an explosion.  But it
may be that the sensitivity of the current simulations occurs only
because theorists are missing essential aspects of physics in the
core.  For example, rotation in the core (Fryer \& Heger 1998; Akiyama
et al. 2003; Kotake, Yamada, \& Sato 2003a; Fryer \& Warren 2004),
asymmetric collapse (Burrows \& Hayes 1996; Fryer 2004a), one-sided
convection due to oscillations (Blondin, Mezzacappa \& DeMarino 2003),
or the merger of modes (Scheck et al. 2003) may help or hinder the
explosion.

The reason we know so little about the core and the resultant
supernova explosion is because we have very few ways of directly
observing the engine behind these explosions.  The collapsed core is
enshrouded by the outer layers of the star and it is not until the
supernova shock reaches the edge of the star and becomes optically
thin to photons that we can finally see the supernova.  Then begins
the laborious task of working back from the optical display to study
the workings of the inner core.  Although these observations (see
Akiyama et al. 2003 and references therein), and observations of the
neutron star remnant (see Lai, Chernoff, \& Cordes 2001; Brisken et
al. 2003 for reviews), have shown that the supernova explosion is
almost certainly asymmetric, the level of asymmetry in the engine is
very difficult to determine from these indirect methods.
Gravitational waves and neutrinos provide the only means to directly
probe the supernova engine.

Neutrinos have only been detected in SN 1987A (Hirata et al. 1987;
Bionta et al. 1987), and gravitational waves (GWs) have yet to be
conclusively detected in supernovae.  Unfortunately, the signals of
both are sufficiently weak to limit their detection to Galactic or
near-Galactic supernovae (Neutrinos: Burrows, Klein \& Gandhi 1998;
GWs: Fryer, Holz \& Hughes: FHH 2002; Dimmelmeier, Font \& M\"uller
2002; Kotake, Yamada \& Sato 2003b; Fryer \& Warren 2004; M\"uller et
al. 2004).  Although predictions for the GW signal from stellar
collapse have been much higher in the past (see FHH 2002; New 2003 for
reviews), these predictions were based on stellar cores rotating much
faster than even the fastest rotating supernova progenitors produced
today (Heger, Langer \& Woosley 2000).  Modern progenitors of stellar
collapse are only marginally, if at all, unstable to bar
instabilities, and strong bar modes (the source of strong
gravitational wave signals) do not occur in collapse calculations
using these progenitors (Fryer \& Warren 2004).  Even if bars did
form, they form at low densities, and at most produce a marginally
detectable signal at 10\,Mpc (Rampp, M\"uller \& Ruffert 1998; FHH
2002).  GW observations, like neutrinos, are limited to Galactic or
near-Galactic supernovae.

Even though the rate of such nearby SNe are low (0.02 year$^{-1}$),
the insight gained from these supernovae will provide essential
information into stellar collapse.  In this paper, we will focus on 3
phases of gravitational waves:
\begin{itemize}
\item[I)] {\bf Stellar Bounce} - The signal produced during the 
bounce caused when the stellar core reaches nuclear densities and 
its runaway free-fall is abruptly halted.  For initial progenitors 
with asymmetries or rotation, the quadrupole moment in the progenitors 
will vary wildly during bounce, producing a strong signal.
\item[II)] {\bf Convection} - The signal produced during the convective 
phase of the supernova engine.  Even if the initial conditions are 
symmetric, the convection modes may merge to form low-mode convection 
with a significant time-variable quadrupole moment.
\item[III)] {\bf Neutrinos} - The signal produced by asymmetries in 
the neutrino emission.  Asymmetries in the collapse or in the convection 
will lead to asymmetries in the neutrino emission.  These asymmetries 
lead to a time-variable quadrupole moment.
\end{itemize}
The GW signal can not only be used to distinguish these phases, but 
also the asymmetry that causes the signal in each phase.

In this paper, we will use the results from recent 3-dimensional
collapse simulations (Fryer \& Warren 2002, 2004; Fryer 2004a) to study
the different characteristics of these signals, focusing especially on
what these signals can tell us about stellar collapse and supernova
explosions.  But a lot of uncertainty remains in such simulations and
we use comparisons with other work and analytic estimates to gauge the
uncertainties in these calculations.  The models, and the numerical
methods used to calculate the GW signal, are discussed in \S 2.  \S 3
shows the results of these simulations, along with comparisons to
other work and analytic estimates.  To make GWs a powerful probe of
the inner core, we must study a bigger picture and correlate the GW
signal with other supernova observables: neutrino signal,
nucleosynthetic yields and other explosion asymmetry characteristics
in the observations (\S 4).  We conclude with a discussion of the
observability of gamma-ray bursts under the collapsar model and a
review of the key observations and the constraints they can place on
the supernova engine.

\section{Computations}

The gravitational wave signals presented here are derived from the
3-dimensional core-collapse calculations by Fryer \& Warren
(2002,2004) and Fryer (2004a).  This core-collapse code
couples smooth particle hydrodynamics (SPH) with a 3-flavor ($\nu_e,
\bar{\nu}_e, \nu_{\mu,\tau}$) flux-limited diffusion neutrino
transport scheme (Warren, Rockefeller, \& Fryer 2004).  Beyond a
trapping radius (corresponding to an optical depth, $\tau_\nu=0.03$),
the neutrinos are assumed to escape the star completely.  The
neutrinos leaking from the ``boundary particles'' at this radius
dominate the observed neutrino flux.  Unless otherwise specified, the
gravity is Newtonian, using the tree-based algorithm described in
Warren \& Salmon (1995).  These models begin with initial stars that
are both rotating and non-rotating, with density distributions that
are either initially symmetric, or perturbed by global asymmetries
(Table 1).

The advantage of such a code is that we can model the collapse of the
entire star in 3-dimensions for a series of initial conditions with
existing computational resources (a typical simulation takes roughly
100,000 processor hours on both the Space Simulator and ASCI Q
machines).  Because the full 3-dimensional star is modeled, no
approximations need be taken to calculate the mass quadrupole moments
and its derivatives.  And since this code is Lagrangian (particle
based), the resolution follows the proto-neutron star and this code is
ideally suited to model asymmetries that accelerate this central core.

However, bear in mind that the neutrinos are modeled with a
single-energy flux-limited transport algorithm.  Comparison of such a
transport algorithm with more sophisticated transport algorithms in 1
and 2-dimensions have found that the such calculations can drastically
change the fate of the collapsing star (Buras et al. 2003).  Note also that
these models assume Newtonian rather than General Relativistic
gravity, which not only affects the bounce of and convective motions
in the core, but also forces us to calculate the gravitational wave
signal using a post-process technique.  All of our results must be 
tempered by these uncertainties.

However, as we shall see in \S 3, the gravitational wave signal is so
weak that the back-reaction of the gravitational wave emission is
unlikely to effect the dynamics in the core.  The remaining
uncertainties are less easy to quantify, but by comparing to 
more detailed 2-dimensional work (M\"uller et al. 2004), we 
can estimate the level of uncertainty in our results (\S 3).

We calculate the gravitational wave signal from baryonic mass motions 
using the formulism of Centrella \& McMillan (1993):
\begin{eqnarray}
c^8/G^2<(rh_+)^2> &=& \frac{4}{15}(\ddot{I}_{xx}-\ddot{I}_{zz})^2+
\frac{4}{15}(\ddot{I}_{yy}-\ddot{I}_{zz})^2+
\frac{1}{10}(\ddot{I}_{xx}-\ddot{I}_{yy})^2 \nonumber \\
&+& \frac{14}{15}(\ddot{I}_{xy})^2 + \frac{4}{15}(\ddot{I}_{xz})^2
+ \frac{4}{15}(\ddot{I}_{yz})^2,
\end{eqnarray}
\begin{equation}
c^8/G^2<(r h_{\times})^2> = \frac{1}{6}(\ddot{I}_{xx}-\ddot{I}_{yy})^2+
\frac{2}{3}(\ddot{I}_{xy})^2+\frac{4}{3}(\ddot{I}_{xz})^2+
\frac{4}{3}(\ddot{I}_{yz})^2,
\end{equation}
where $r$ is the distance to the source.  The angle brackets in these
equations denote averaging over all source orientations, so that for
example
\begin{equation}
\langle (r h_+)^2 \rangle = {1\over4\pi} \int d\Omega r^2
h_+(\theta,\phi)^2\;.
\label{eq:averaging}
\end{equation}
The quantities ${\ddot I}_{ij}$ are the second time derivatives of the
trace-free quadrupole moment of the source.  The diagonal elements are
given by, for example,
\begin{equation}
{\ddot I}_{xx} = {2\over3}\sum_{p = 1}^N m_p(2 x_p {\ddot x}_p - y_p
{\ddot y}_p - z_p {\ddot z}_p + 2 {\dot x}_p^2 - {\dot y}_p^2 - {\dot
z}_p^2),
\label{eq:diag_quadrupole}
\end{equation}
where $m_p$ is the mass of an SPH particle, and $(x_p,y_p,z_p)$ is its
coordinate location.  The other diagonal elements can be obtained
using Eq.\ (\ref{eq:diag_quadrupole}) plus cyclic permutation of the
coordinate labels: ${\ddot I}_{xx} \to {\ddot I}_{yy}$ via $x\to y$,
$y \to z$, $z \to x$; ${\ddot I}_{xx} \to {\ddot I}_{zz}$ via $x
\to z$, $y\to x$, $z \to y$.  The off-diagonal elements are given by,
for example
\begin{equation}
{\ddot I}_{xy} = \sum_{p = 1}^N m_p(x_p {\ddot x}_p + y_p
{\ddot y}_p + 2 {\dot x}_p{\dot y}_p),
\label{eq:offdiag_quadrupole}
\end{equation}
the remaining off-diagonal elements can be found via symmetry (${\ddot
  I}_{ji} = {\ddot I}_{ij}$) plus cyclic permutation.  This approach
provides only averaged square amplitudes, instead of the specific
amplitude of the gravitational wave signal.  This paper is concerned
with general qualitative features of the wave signals, rather then
detailed predictions. In this vein, the averaged square amplitudes are
a cleaner and more effective description.

For the trends we a
studying here, this formulism is sufficient and provides a cleaner
signal.

For gravitational waves from neutrinos, we use the formulae derived 
by M\"uller \& Janka (1997):
\begin{eqnarray}
(h^{\rm TT}_{xx})_{\rm pole} &=& \frac{2G}{c^4 r} \int_{- \infty}^{t-R/c} 
dt' \nonumber \\
& & \int_{4 \pi} d \Omega ' (1+cos \theta ') cos (2 \phi ') \frac{dL_\nu 
(\Omega ', t ')}{d \Omega '},
\end{eqnarray}
\begin{eqnarray}
(h^{\rm TT}_{xx})_{\rm equator} &=& \frac{2G}{c^4 r} \int_{- \infty}^{t-R/c} 
dt' \int_{4 \pi} d \Omega ' (1+sin \theta ' cos \phi ') \nonumber \\
& & \frac{cos^2 \theta ' - sin^2 \theta ' sin^2 \phi '} 
{cos^2 \theta ' + sin^2 \theta ' sin^2 \phi '}
\frac{dL_\nu (\Omega ', t ')}{d \Omega '}.
\end{eqnarray}
where $G$ is the gravitational constant, $c$ is the speed of light and
$r$ is the distance of the object.  $(h^{\rm TT}_{xx})_{\rm pole}$
denotes the emergent strain for an observer situated along the source
coordinate frame's z-axis (or pole) and $(h^{\rm TT}_{xx})_{\rm
  equator}$ is the comparable strain for an observer situated
perpindicular to this z-axis (or equator).  For our simulations, the
rotation axis or asymmetry axis is the z-axis and the polar view
corresponds to the axis of symmetry for the initial perturbations.

In SPH calculations, these expressions reduce to:
\begin{equation}
(h^{\rm TT}_{xx})_{\rm pole} = \frac{2G}{c^4 R} \sum \Delta t 
\sum_{p=1}^{N} (1 \pm z/r) (2 x^2/r^2 - 1) \Delta L_{\nu},
\end{equation}
\begin{equation}
(h^{\rm TT}_{xx})_{\rm equator} = \frac{2G}{c^4 R} \sum \Delta t 
\sum_{p=1}^{N} (1 \pm \sqrt{1-z^2/r^2} (x,y)/r) 
\frac{z^2/r^2 - \sqrt{1-z^2/r^2}(x,y)^2/r^2}
{z^2/r^2 + \sqrt{1-z^2/r^2}(x,y)^2/r^2} \Delta L_{\nu}.
\end{equation}
where $\Delta t$ is the timestep, the summation is over all particles
emitting neutrinos that escape this star (primarily, the boundary
particles).  We assume that the neutrino emission at these boundaries
is along the radial direction of the particle, so $\Delta L_{\nu}$ is
set to $E_\nu/\Delta t$ where $E_\nu$ is the neutrino energy emitted
by that particle.  If the neutrinos are more isotropic, this will
weaken the gravitational wave signal, but not by more than a factor of
2.  The $\pm$ correspond to observers along the positive/negative
directions of each axis: for instance, in the polar equation, this
corresponds to the positive/negative z-axis.  In the equatorial 
region, the x,y axis is determined by the choice of x or y position 
for the (x,y) coordinate in the equation.

\section{Results from 3-dimensional Simulations}

A massive star ends its life when its iron core becomes so massive
that that the thermal and electron degeneracy pressure of the star can
no longer support the core and it implodes.  This implosion continues
until the core reaches nuclear densities, where nuclear forces and
neutron degeneracy pressure halt the collapse.  The ``bounce'' of the
core does not produce an explosion.  The bounce shock stalls when
neutrino cooling and photodisintegration saps the shock of its energy.
If there are asymmetries in the star, this collapse and bounce phase
produces rapidly varying quadrupole moments and gravitational waves.

The stalled shock leaves behind a convectively unstable region between 
the proto-neutron star and the accretion shock of the infalling star.  
Neutrinos leaking out of the proto-neutron star heat this region, 
driving further convection.  When the pressure from this convection 
overcomes the pressure of the infalling star, a supernova explosion 
is launched.  The mass motions in this convection also produce 
rapidly varying quadrupole moments.

If stellar asymmetries or convective motions are sufficiently
asymmetric, the neutrino emission will also be asymmetric.  These 
neutrino asymmetries can produce a significant gravitational 
wave signal.  As we shall see, these signals can dominate the 
GW signal in some cases.

Let's take a closer look at each of these phases in turn, using 
core-collapse simulations as a guide.

\subsection{Gravitational Waves from Core Bounce}

{\bf Rotation:} Although massive stars are observed to rotate at
nearly break-up velocities while on the main-sequence, the angular
velocity (of the envelope at least) decreases dramatically (just from
angular momentum conservation) when these stars expand into giants.
If they cores are not coupled to their expanding envelopes, they will
spin up as the core contracts.  If they are coupled, the contracting
core will impart its angular momentum to the envelope and slow down as
this envelope expands.  The extent of this coupling depends
sensitively on uncertain magnetohydrodynamic physics and current
supernova models (Heger, Langer, \& Woosley 2000; Heger, Woosley, \&
Spruit 2004) predict a range of answers (see Fryer \& Warren 2004 for
a review).  If these cores collapse with significant angular velocity,
centrifugal suppport will deform the collapse, making strong
asymmetries and, ultimately, a strong gravitational wave signal.  GW
observations provide an ideal means to probe this asymmetry.

Figure 1 shows the gravitational wave signals for a range of rotating
stellar collapses.  Model Rot1 is the the fastest spinning core
produced by Heger, Langer \& Woosley (2000).  The signal in Fig. 1
assumes the collapse occurs 10\,kpc away.  At 10\,kpc (within the
Galaxy), this bounce signal from models Rot1 and Rot2 (peaking at
$2\times 10^{-21}$ at $\sim$1\,kHz) is well above the expected noise 
level of advanced LIGO, suggesting it should not be difficult to 
observe.  The strong signal dies away very quickly and is strongest right
at core bounce.  But the strength of this signal depends sensitively
on the rotation.  The slower rotating cores do not produce
gravitational wave signals (Rot3) that are significantly stronger than
those produced by asymmetries (Rot4 and Rot5)\footnote{Our SPH
  initial conditions introduce $\sim$5-10\% asymmetries.  Although
  these are numerical artifacts, they do not differ significantly from
  the asymmetries expected from explosive burning prior to collapse.}
in a non-rotating core.

{\bf Asymmetries from Explosive Burning:} Rotation is not the only way
to produce asymmetries in the star just prior to collapse.  Explosive
burning in the oxygen and silicon shells produces strong convection in
these shells, causing density perturbations in these shells (Bazan \&
Arnett 1998).  If this convection has many downflows and upflows (high
mode convection), these density perturbations would be randomly
distributed across the oxygen and silicon layers of the star, similar
to the 5-10\% perturbations introduced using the smooth particle
hydrodynamics technique.  The signal from our non-rotating star (Figs.
1,2) is caused by these pertubations.

But what if the convection is dominated by low modes with the density
perturbations taking on a more global scale?  Burrows \& Hayes (1996)
argued that global perturbations from oxygen and silicon burning would
lead to asymmetries in the bounce and could be the source of neutron
star kicks.  Lai \& Goldreich (2000) argued that this convection would
drive oscillatory modes in the iron core that would grow during
collapse and drive large asymmetries in the bounce.  Fryer (2004a)
mimicked these global perturbations by decreasing the density to the
oxygen and silicon layers (or in the entire star) in a cone an the
positive z axis.

Fig.  3 shows the gravitational wave signals from these globabally
asymmetric explosions.  The bounce signal from these global
asymmetries is much different than that of the rotating stars.  The
peak signal does not occur at bounce, but 10-20\,ms later.  The signal
from baryonic mass motions is dominated by the asymmetries caused by
the oscillations in the core as ejecta and neutrino asymmetries kick
the neutron star.  

Models Asym1, Asym2 correspond to the large asymmetry cases for
``Shell Only'', or ``With Oscillations'' simulations in Fryer (2004a).
Models Asym3, Asym4 are 30\% perturbations in the oxygen/silicon
burning shells.  Asym3 and Asym4 differ in that the momentum imparted
by the last scattering of neutrinos is included in Asym3 and not in
Asym4\footnote{The flux limited-diffusion scheme used by Fryer (2004a)
  includes the effects of neutrino pressure and momentum.  Below
  $\tau_{\nu}=0.03$, we make a light-bulb approximation.  At the
  boundary, the neutrino momentum must be included.  The prescription
  for this is described in Fryer (2004a).}.  Note that there is very
little difference between the GW signal of these two models.  Neither
model produces sizable kicks ($v_{\rm kick}<30$km\,s$^{-1}$) in the
neutron star or strong oscillations in the core.  Models Asym1 and
Asym2 produce much stronger oscillations and slightly stronger GW
signals.  The largest signal occurs when oscillatory modes are driven
in the iron core prior to collapse (Asym1).  However, as we shall see
in \S \ref{sec:neutrino}, the neutrino GW signal for these models is
much stronger than that of mass motions, and that will dominate the
signal at late times.

Fig. 4 compares the gravitational wave signals from the extreme 
cases of rotating (Rot1) and globally asymmetric (Asym1) simulations 
along with a non-rotating model (SPH1 - equivalent to high mode 
perturbations).  For mass motions, the rotating model has the strongest 
signal which peaks at bounce.  The globally asymmetric model can 
also produce a signal which should detectable by LIGO, peaking at  
$\sim$10-20\,ms after bounce.  If the rotation or 
asymmetries are not extreme, the gravitational wave signal will 
not be easily detectable.  However, even a non-detection will 
tell us much about the core, as it will rule out these large 
(or larger) asymmetries in the initial core.

\subsection{Gravitational Waves from Convection}

With accurate equations of state, when the bounce shock stalls, it
leaves behind an unstable entropy profile above the proto-neutron
star.  The region bounded by the proto-neutron star on the inside and
the accretion shock where the rest of the material falls down onto the
stalled bounce shock quickly becomes convective.  This convection is
further driven by heating from neutrinos leaking out of the
proto-neutron star.  Convection allows the energy deposited from
neutrinos to convert into kinetic energy of rising matter.  When this
matter is able to push off the infalling star, an explosion is
launched (see Fryer 1999 for a review).

Herant et al. (1994) found that the energy in this convection was
primarily carried in just a few upflows and Herant (1995) argued that
if the convection were dominated by a single upflow, they could
explain the high space velocities of pulsars (see \S \ref{sec:remnant}).
Indeed, the convective modes merge with time (Fig. 5).  In our 
3-dimensional models, the explosion occurs so quickly that there are 
still $\sim$3-6 upflows when the explosion is launched.  Hence, the 
convective modes do not produce a strong signal for our models (Figs. 1-4).
Indeed, the only simulations that produced strong signals from baryonic 
mass motions were the globally asymmetric collapse simulations with 
strongly oscillating neutron stars.

However, Scheck et al. (2003) have found that for sufficiently delayed
explosions (which may well be more typical in supernovae), the
convective upflows can merge until there is only 1 single upflow.
This can produce a gravitational wave signal that rivals even the
bounce signal for some rotating stars (M\"uller et al. 2004).  To get
a better understanding of this signal, we can approximate the mass
motions by assuming that the convective motions have merged, leading
to a single downflow, but that the downflow occurs in spurts (large
blobs accreting).  Such an assumption is not too unreasonable, as mass
builds up at the accretion shock until its weight carries it down,
punching its way through the upflowing material.  To estimate the
maximum signal from convection, we assume the mass of the accreting
blob is 0.1\,M$_\odot$ and it accelerates at free-fall from the edge
of the accretion region to half way between the accretion shock and
the proto-neutron star surface (at 50\,km), at which point they begin
to decelerate and stop at the proto-neutron star surface (we chose a 
deceleration that was symmetric in magnitude about the half radius).  
The signal produced by this ``convection'' varies as the blob falls.  
Figure 6 plots the maximum of that signal as a function of the 
accretion shock radius.  As the convection region expands, the 
signal will decrease, but it could take 500\,ms for the shock to 
move from 100\,km to 500\,km.  Note that this signal is an upper 
limit to the possible signal from convection and is an order of 
magnitude higher than that predicted by M\"uller et al. (2004).  
It is also nearly an order of magnitude greater than the rotating 
simulations presented here.

\subsection{Gravitational Waves from Neutrino Asymmetries}
\label{sec:neutrino}

All of these asymmetries in the bounce and convection lead to asymmetries in
the neutrino emission.  These asymmetries produce a gravitational wave
signal that grows with time.  In most cases, the asymmetry in the
neutrinos is not large enough to dominate the GW signal (e.g. M\"uller
et al. 2004), but neutrino asymmetries can dominate the gravitational 
wave signal in the case of asymmetric collapse (e.g. Burrows \& Hayes; 
Fryer 2004a).  

Figs. 7 and 8 show the gravitational wave signal from models Asym1 and
Asym2 respectively.  The signal continues to grow with time and, for
these two simulations, exceeds the bounce or convective signals by
over an order of magnitude!  Unfortunately, these signals were not
calculated for any models prior to Fryer (2004a), so we can not present
gravitational wave signals for earlier models.  But the neutrino
asymmetry in the rotating and spherical models are more than $\sim 2$
orders of magnitude less than the asymmetries of these simulations,
so neutrinos asymmetries may not play a dominant role in most
supernovae.  Indeed, M\"uller et al. (2004) found that the GW signal from
asymmetric neutrino emission was slightly lower than that produced 
by baryonic mass motions.

Figures 7 and 8 show the GW signal from neutrinos along 6 lines of
site (positive and negative x, y, and z axes).  The magnitude of the
signal is strongest along the positive z axis (the positive z axis is
where the initial density perturbation was placed).  Reviewing all
axes gives a handle on how observation location will effect the
signal.  These extreme asymmetries will easily produce signals
detectable in a Galactic supernova.  In addition, the time evolution
of this signal is significantly different that it can easily be
distinguished from signals by mass motions.  Unfortunately, unlike the
signal from mass motions, the signal for neutrinos peaks at lower
frequencies (Fig. 9) and, in advanced LIGO sensitivity band, the
neutrino asymmetry GW signal is only slightly stronger than mass
motions for these stellar implosions.

Detecting a Galactic supernova in gravitational waves and following
the time evolution of the GW signal can easily tell us much about the
behavior of the inner core of a star during collapse.  We can
distinguish between rotating and asymmetric collapse and between
low-mode and high-mode convection.  We can also estimate the level 
of neutrino asymmetry.  Gravitational waves provide an ideal window 
into the mass motions in, and just above, the proto-neutron star 
core.

\section{Comparing with other Observables}

Especially with Galactic supernovae, we have a number of additional
observations which can be used to help learn about supernovae ranging
from direct observations of the core with neutrinos to indirect
methods such as supernova asymmetries, nucleosynthetic yields, and
studies of the compact remnant from these explosions.  Combined with 
these constraints, gravitational waves can tell us much about the 
supernova engine.

\subsection{Neutrinos} 

Neutrino detections provide the only other means beyond GWs to study
the supernova engine directly.  SuperKamiokande will detect over 5000
neutrinos from a supernova 10\,kpc away (Burrows, Klein, \& Gandhi
1998).  The bulk of these detections will be electron anti-neutrinos,
but both SuperKamiokande and the Sudbury Neutrino Observatory will
detect a few hundred electron and $\mu$ neutrinos.  This signal will
be sufficient to produce reasonable neutrino light-curves that can be
used to make detailed comparisons with the neutrino emission predicted
by models.

How do the mass motions affect the neutrino observations?  The actual
net asymmetries in the neutrino emission tend to be small (a percent
or less - see Janka \& M\"onchmeyer 1989a,1989b), but Kotake et
al. (2003a) have found that rotation can lead to a neutrino energy
that varies by as much as a factor of 2 for different angular
lines-of-sight.  Figure 10 shows the electron neutrino luminosity and
energy for a range of models (from spherical to rotating to asymmetric
collapse).  For our models, the electron neutrino luminosity does not
change much even though the actual progenitors for these objects does
vary in some cases.  The electron neutrino energies vary by $<$30\%.
However, bear in mind that these calculations rely upon a
single-energy flux-limited diffusion scheme and we should take these
quantitative results with a grain of salt.

The $\mu$, $\tau$, and electron anti-neutrinos vary much more than the
electron neutrinos (Figs. 11-12). Even so, it would be very
difficult to distinguish asymmetries in the core with neutrinos.  But
neutrinos do give us an ideal probe into the equation of state in the
core (e.g.  Pons et al.  2000).  The details of the equation of state
will also affect the gravitational waves to a lesser extent
(Dimmelmeier et al. 2002) and we can use neutrino observations to
distinguish equation of state effects from mass motions.  Combining
the neutrino and the gravitational wave signal, we can study both the
mass asymmetries in the collapse and the behavior of matter at nuclear
densities.  Quantitative analyses will require much more detailed
core-collapse models.

The detection of both gravitational waves and neutrinos also has
implications for calculating the neutrino mass.  The delay between the
emission of neutrinos ($t_e$) and their arrival at a detector
($t_d$)is (e.g.  Arnett \& Rosner 1987):
\begin{equation}
t_d-t_e \approx (d/c) (1+0.5 m_\nu^2/E_\nu^2)
\end{equation}
where $d$ is the distance from the supernova to the observer, $c$ is
the speed of light and $m_\nu,E_\nu$ are the mass and energy of the
neutrino respectively.  If we know that our GW signal peaks at bounce
(as is the case for our rotating supernovae), the time between the
peak emission of the gravitational and neutrino signals should not
differ by more than 5\,ms and can easily be determined to this
accuracy.  By differencing the delay in the neutrinos by the delay in
the gravitational wave signal $(t_d-t_e)_\nu - (t_d-t_e)_{\rm GW}$, we
could then determine the neutrino mass from a 10\,kpc supernova to
better than 1eV (see also, Arnaud et al. 2002).  It will be more difficult 
to use this technique if the neutrino signal is determined by 
convective modes or neutrino asymmetries, where the timing is less 
accurate.

\subsection{Observations of Explosion Asymmetries}
The same asymmetric mass motions that produce gravitational waves may
also produce asymmetries in the supernova explosion.  For example,
rotation causes the strongest convection to occur along the rotation
axis and this bipolar convection will drive asymmetric explosions
(Fryer \& Heger 2000; Kotake et al. 2003a; Fryer \& Warren
2004).  The merger of convective modes will also produce wildly
asymmetric supernovae (Scheck et al. 2003).  Burrows \& Hayes (1996)
also argued that asymmetric collapse would produce asymmetries in the
explosion, but 3-dimensional calculations have found that these
asymmetries are mild (Fryer 2004a).

For Galactic supernovae, there are many observational diagnostics that
can help, albeit indirectly, determine the level of asymmetry in the
core.  Supernova 1987A provides a number of examples of how powerful
these diagnostics can be for nearby supernovae and modern telescopes.
One of the surprises SN 1987A provided for astronomers was that
somehow the nickel in the core was mixed well into the star, causing
the gamma-ray luminosity powered by the $^{56}$Ni produced in this
explosion to peak 150\,d earlier than expected by theorists (Pinto \&
Woosley 1988).  Spherical explosions can not explain this easily
(Kifonidis et al. 2003), but mild ($\sim$ 2 times stronger along the
rotation axis over the equator) asymmetries would easily explain this
mixing (Fig. 13: Hungerford, Fryer, \& Warren 2003, Fryer 2004b).
Because of these asymmetries, the nickel is also ejected
preferentially along the asymmetry axis (Fig. 14) and this asymmetric
distribution leads to a gamma-ray line profile that, in principal, can
determine the angle and level of the asymmetry (with some uncertainty
due to the degeneracy of these two effects: Hungerford et al. 2003).
Supernova 1987A was too distant to easily make these distinctions, but
a Galactic supernova with modern telescopes would provide strong
constraints.

Asymmetries also effect the nucleosynthetic yields, optical line
profiles, and polarization of supernovae.  Varying the shock velocity
changes the yields from explosive nuclear burning.  Nagataki et
al. (1998) found that mild asymmetries (on the same level as those
required to explain the gamma-ray lines) were required to best fit the
nucleosynthetic yields of SN 1987A.  Interpretations of the
polarization signal have argued that most supernovae must be jet-like
with explosions 100 times stronger along the rotation axis over the
equatorial plane (Akiyama et al. 2003).  However, this interpretation
requires several layers of detailed radiation transport calculations.
This result is in contrast to more direct measurements from both radio
studies of supernovae (Berger et al. 2003) and specific studies of SN
1987A (e.g. the mixing and nucleosynthetic yield results described
above) which argue for mild asymmetries.  It is likely that the
interpretation of the polarization measurement is overestimating the
asymmetry and such large asymmetries are probably limited to a small
fraction of all supernovae.  But this descrepancy highlights the
limitations of such indirect measurements of asymmetries.

What indirect observations from current supernovae can tell us at this
point is that asymmetries exist.  The situation will change with a
Galactic supernova.  A Galactic supernova will provide enough
gamma-ray photons (which have few transport uncertainties) to
constrain not only the level of asymmetry, but the axis of asymmetry.
Such information can be used to help us understand the GW signal and
use gravitational waves to constrain the mass motions.

\subsection{Compact Remnants}
\label{sec:remnant} 
Asymmetric mass motions also affect the velocity and spin of the
neutron star remnant produced in the supernova explosion.  Spinning,
magnetized neutron stars are observed as pulsars.  For the fastest
spinning collapse progenitors (the ones that produce sizable GW
signals), the angular momentum is so high in the core that, if no
angular momentum were lost in the supernova explosion, the resultant
core would be spinning at sub-millisecond periods.  During the
collapse and subsequent supernova explosion of the core, the high
angular momentum in the core is ejected, but the newly-born neutron
star will have periods in the 1-3\,ms range, with total rotational
energies in excess of $10^{51}$\,ergs (Fryer \& Warren 2004).  If such
a pulsar had moderate to high (above $10^{11}$\,G) magnetic fields,
its emission would easily be observable in the Galaxy (even if not
directed toward us!).  Indeed, the pulsar emission may alter the
supernova explosion energy (see Fryer \& Warren 2004). A detailed,
strong GW signal (or the lack of a signal) could give constraints on
this emission.  Similarly, pulsar observations can help interpret the
GW signal.  Since the pulsar spin depends primarily on the rotation of
the core, it is most sensitive to this source of gravitational waves.
A number of limitations may confuse such results: e.g. the neutron
star could lose angular momentum due to GW driven modes after the
explosion, some collapsing stars produce fast-spinning neutron stars
but only weak GW signals.

The evidence that neutron stars receive a sizable ``kick'' ($\sim
500$\,km\,s$^{-1}$) at birth continues to grow (see Lai, Chernoff, \&
Cordes (2001); Brisken et al. (2003) for recent reviews).  These kicks
are either produced by ejecta asymmetries (e.g. Herant 1995; Burrows
\& Hayes 1996; Scheck et al. 2003) or neutrino asymmetries (Lai \&
Arras 1999; Fryer 2004a).  It is likely that the velocity of the
neutron star remnant of a Galactic supernova will be measured.  The
velocity of the neutron star does not depend so much on the rotation,
but may depend on the level of asymmetry in the collapse (Burrows \&
Hayes 1996; Fryer 2004a) or the merger of convective modes (Herant
1995; Scheck et al. 2003).  Because these two GW mechanisms have such
distinct signatures, GW observations of a Galactic supernova (combined
with a neutron star velocity measurement) will be able to determine
which, if either, of these mechanisms produce neutron star kicks.

\section{Conclusions}

A number of asymmetries in the stellar collapse can produce a
gravitational wave signal in a Galactic supernova that should be
detectable by advanced LIGO.  These asymmetries are caused by rotation
in the star, asymmetries induced by explosive nuclear burning just
prior to collapse, or low mode convection in the supernova engine.
Each has a distinctive gravitational wave signal which can be used to
study the mass motions in the supernova engine.

But the real power of GWs arises when combined and corroborated with
other observations.  The correlation between GW and these observations
is summarized in Tables 2 and 3.  Fast rotating progenitors can produce a
strong signal at bounce, mild asymmetries in the ejecta and
fast-spinning neutron star remnants.  The timing of the GW signal is
within 5\,ms (maybe even closer) of the neutrino signal and such an
event could be used to constrain the neutrino mass.  Asymmetries in
the collapsing star caused by explosive burning can produce strong
signals at late times through asymmetric neutrino emission and mild
velocities on the neutron star, but the explosion is roughly
symmetric.  Low mode convection will produce a gravitational wave
signal during the convective engine phase, asymmetric explosions and
possibly strong neutron star kicks.  The GW signal will help determine 
the mechanism behind the SN and our understanding of these additional 
phenomena.

Supernovae are not the only explosions produced by stellar collapse.
The favored mechanism for long-duration gamma-ray bursts, the
collapsar model (Woosley 1993; MacFadyen \& Woosley 1999), invokes the
collapse of a massive star.  Collapsars are massive stars that do not
produce strong supernova explosions, but instead collapse to form
black holes.  If the star is rotating fast enough, the high angular
momentum stellar material will form an accretion disk around the black
hole.  Energy liberated from the disk drives a strong explosion (even
stronger than supernovae) in a jet along the rotation axis.

Because rotation is required to produce a collapsar explosion, the
collapsing stars will necessarily be fast rotators.  Thus they will
have a strong bounce GW signal.  However, this signal is unlikely to
be much stronger than our fast rotating collapse models (too much
rotation will also prevent the collapsar engine from working -
MacFadyen \& Woosley 1999).  It is likely that collapsars arise from
extremely massive stars (Heger et al. 2003), but, because the cores of
massive stars are essentially all the same, this does not alter the GW
signal significantly from our extreme rotating case (Dupuis, Fryer, \&
Heger 2004).  

The bounce of collapsar driven gamma-ray bursts will only be
detectable when they occur in the galaxy.  The accretion onto the
black hole will cause ringing, but this signal too will only be
detectable in a Galactic gamma-ray burst.  Finally, for the low-mass
disks produced in collapsars (Popham, Fryer, \& Woosley 1998;
MacFadyen \& Woosley 1999), no disk instabilities will form and the
accretion phase of collapsars are unlikely to produce strong GW
signals.  The narrowly collimated, low-baryon jet produced in
collapsars will produce a signal akin to neutrino driven signals.
However, for a bipolar jet, there will be no signal along the jet
axis.  Perpendicular to the jet axis, the signal (using equation 8)
can be as high as $5 \times 10^{-24}$ for a $10^{51}$\,erg jet, on par
with the signal from the disk, but still only detectable within the
Galaxy and only if the jet is not directed toward us (so not
associated with GRBs, but maybe with hypernovae).  What this all boils
down to is that collapsar driven gamma-ray bursts, like supernovae,
will not be detectable beyond the Galaxy.  Given their low rate ($\sim
10^{-5} {\rm year^{-1}}$ in the Galaxy), collapsars will not be a
strong GW source.  However, other gamma-ray burst models will produce
strong GW signals (e.g. neutron star mergers) and GWs can easily
distinguish these two burst engines.

\acknowledgements

It is a pleasure to thank L. S. Finn whose questions set up the
direction of this paper.  This work under the auspices of the
University of Arizona and the U.S. Dept. of Energy, and supported by
its contract W-7405-ENG-36 to Los Alamos National Laboratory as well
as DOE SciDAC grant number DE-FC02-01ER41176, NSF grant PHY-0244424
and NSF Grant PHY-0114422 to the CfCP.  The simulations used in this
work were run on LANL's ASCI Q machine and the Space Simulator.

\clearpage
\begin{deluxetable}{llccc}
  \tablewidth{38pc} 
\tablecaption{Collapse Models} 

\tablehead{ \colhead{Model} & \colhead{Reference} 
& \colhead{Initial Rot.} & \colhead{Initial} 
& \colhead{Number of}\\
  \colhead{Name} & \colhead{Name}  & \colhead{(rad\,s$^{-1}$)} 
& \colhead{Asymmetry} & \colhead{Particles}}

\startdata

Rot1 & SN15A-hr\tablenotemark{a} & 4 & symmetric & $5 \times 10^{6}$ \\
Rot2 & SN15B\tablenotemark{a} & 10 & symmetric & $5 \times 10^{5}$ \\
Rot3 & SN15C\tablenotemark{a} & 0.25 & symmetric & $10^{6}$ \\
Rot4 & SN15B-nr\tablenotemark{a} & 0 & symmetric & $10^{6}$ \\
Rot5 & SN15A-nr\tablenotemark{a} & 0 & symmetric & $10^{6}$ \\
Sph1 & Model C\tablenotemark{b} & 0 & symmetric & $3 \times 10^{6}$ \\
Sph2 & Model B\tablenotemark{b} & 0 & symmetric & $10^{6}$ \\
Sph3 & Model A\tablenotemark{b} & 0 & symmetric & $3 \times 10^{5}$ \\
Asym1 & Shell Only\tablenotemark{c} & 0 & 40\% in O,Si\tablenotemark{d} & $10^{6}$ \\
Asym2 & With Oscillations\tablenotemark{c} & 0 & 25\% in Fe,O,Si\tablenotemark{d} & $10^{6}$ \\
Asym3 & Shell Only\tablenotemark{e} & 0 & 30\% in O,Si\tablenotemark{d} & $10^{6}$ \\
Asym4 & Shell Only\tablenotemark{e} & 0 & 30\% in O,Si\tablenotemark{d,f} & $10^{6}$ \\

\enddata

\tablenotetext{a}{Fryer \& Warren (2004)} 
\tablenotetext{b}{Fryer \& Warren (2002)} 
\tablenotetext{c}{Fryer (2004a)}
\tablenotetext{d}{This refers to the density decrease in a
  $30^{\circ}$ cone along the positive z axis either in just the O, Si
  layers (Shell Only) or the entire star (With Oscillations).}
\tablenotetext{e}{Simulations first presented here, but using the same
  conditions as in Fryer (2004a).}  
\tablenotetext{f}{Backreaction from momentum carried away by neutrinos
  not included.}

\end{deluxetable}

\begin{deluxetable}{lccccc}
  \tablewidth{38pc} 
\tablecaption{Collapse Results} 

\tablehead{ \colhead{Model} & \colhead{Peak GW Signal} & 
\colhead{f(Hz)} &
\colhead{Pulsar} & \colhead{NS Velocity} & 
\colhead{Asymmetry} \\
\colhead{Name} & \colhead{($10^{-21}$ at 10\,kpc)} & 
\colhead{at Peak\tablenotemark{a}} & \colhead{Period} & 
  \colhead{km\,s$^{-1}$} & \colhead{$v_{\rm pole}$/$v_{\rm eq}$}}

\startdata

Rot1 & 2.3 & $\sim$1000 & $>0.66$\,ms & $<30$ & $\sim 2$ \\
Rot2 & 1.9 & $\sim$1000 & $>0.35$\,ms & $<30$ & $\sim 1.5$ \\
Rot3 & 0.5 & $\sim$1000 & $>17$\,ms & $<30$ & $<1.1$ \\
Rot4 & 0.4 & $\sim$100  & $>1$\,s & $<30$ & $<1.1$ \\
Rot5 & 0.5 & $\sim$200  & $>1$\,s & $<30$ & $<1.1$ \\
Sph1 & 0.1 & $\sim$200  & $>1$\,s & $<30$ & $<1.1$ \\
Sph2 & 0.1 & $\sim$100  & $>1$\,s & $<30$ & $<1.1$ \\
Sph3 & 0.1 & $\sim$100 & $>1$\,s & $<30$ & $<1.1$ \\
Asym1 & 0.9 & $\sim$1000\tablenotemark{b} & $>1$\,s & $\sim200$ & $\sim 1.2$ \\
Asym2 & 0.4 & $\sim$1000\tablenotemark{b} & $>1$\,s & $<100$ & $\sim 1.2$ \\
Asym3 & 0.2 & $\sim$1000\tablenotemark{b} & $>1$\,s & $<30$ & $<1.1$ \\
Asym4 & 0.3 & $\sim$1000\tablenotemark{b} & $>1$\,s & $<30$ & $<1.1$ \\

\enddata

\tablenotetext{a}{Because most of the signals are dominated by a single
  burst, this frequency is just $t_{\rm burst}^{-1}$ where $t_{\rm
    burst}$ is the burst duration.}
\tablenotetext{b}{This is the frequency of the baryonic mass motions.  
  Neutrino asymmetries dominate the signal below this frequency, peaking 
  below 10 Hz.}

\end{deluxetable}

\begin{deluxetable}{lcccccc}
  \tablewidth{42pc} 
\tablecaption{GW sources in Core-Collapse} 

\tablehead{ \colhead{Initial} &  \multicolumn{3}{c}{Gravitational Waves} & \multicolumn{3}{c}{Correlation} \\
\colhead{Asymmetry}  & \colhead{Source} & \colhead{Peak GW Signal\tablenotemark{a}} & \colhead{$T_\nu-T_{\rm GW}$} & \colhead{Expl. Asym.} & \colhead{ms Pulsar}& \colhead{NS Kick}}

\startdata

None   & Convection\tablenotemark{b} & $10^{-22}$ & 50-500ms & Y? & N & Y? \\
Rotation & Bounce\tablenotemark{b} & $2.5\times10^{-22}$ & 0\,ms & Y & Y & N \\
Density\tablenotemark{c} & Neutrinos & $>4\times10^{-22}$ & 50ms-1s & N & N & Y? \\

\enddata
\tablenotetext{a}{at 10\,kpc}
\tablenotetext{b}{Baryonic mass motions drive signal.}
\tablenotetext{c}{Density Perturbation caused by explosive shell 
burning prior to collapse.}

\end{deluxetable}

\clearpage
\begin{figure}
\plotone{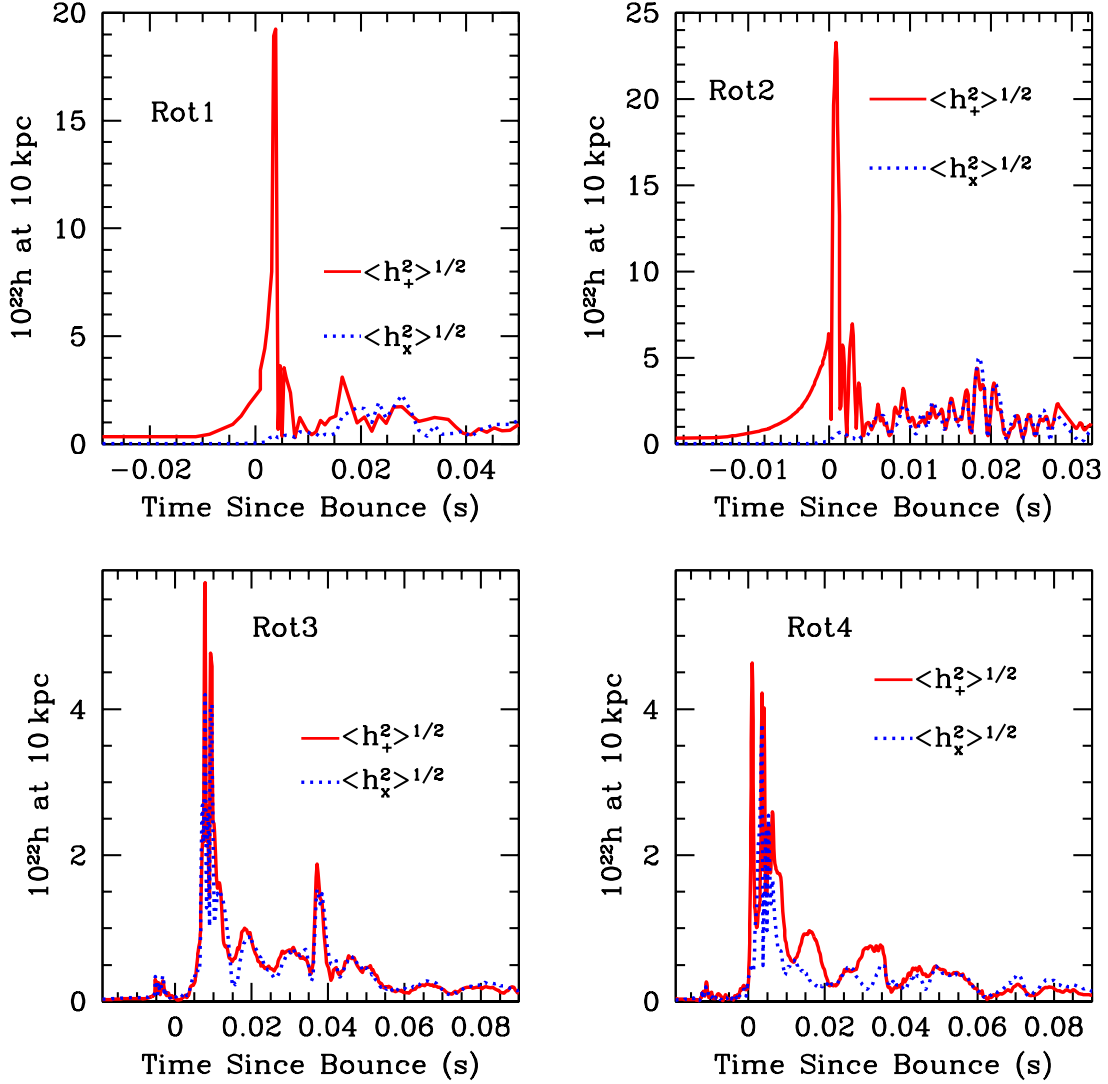}
\caption{The angle averaged wave amplitudes ($<h^2_{+}>^{1/2}$: solid
  line,$<h^2_{x}>^{1/2}$: dotted line) for the mass motions from 4
  rotating supernova models.  Rot2 is the fastest rotating model.
  Rot3 is what would be predicted for a magnetically braked core.
  Rot4 shows the signal from the same star as Rot2, but where the
  rotation was set to zero just before collapse.  See Table 1 for more
  details.  Note that the gravitational wave signal is a factor of 5
  higher in the rapidly versus slowly rotating models.  A fast-rotating
  supernova in the Galaxy should be detectable by advanced LIGO.}
\end{figure}
\clearpage

\begin{figure}
\plotone{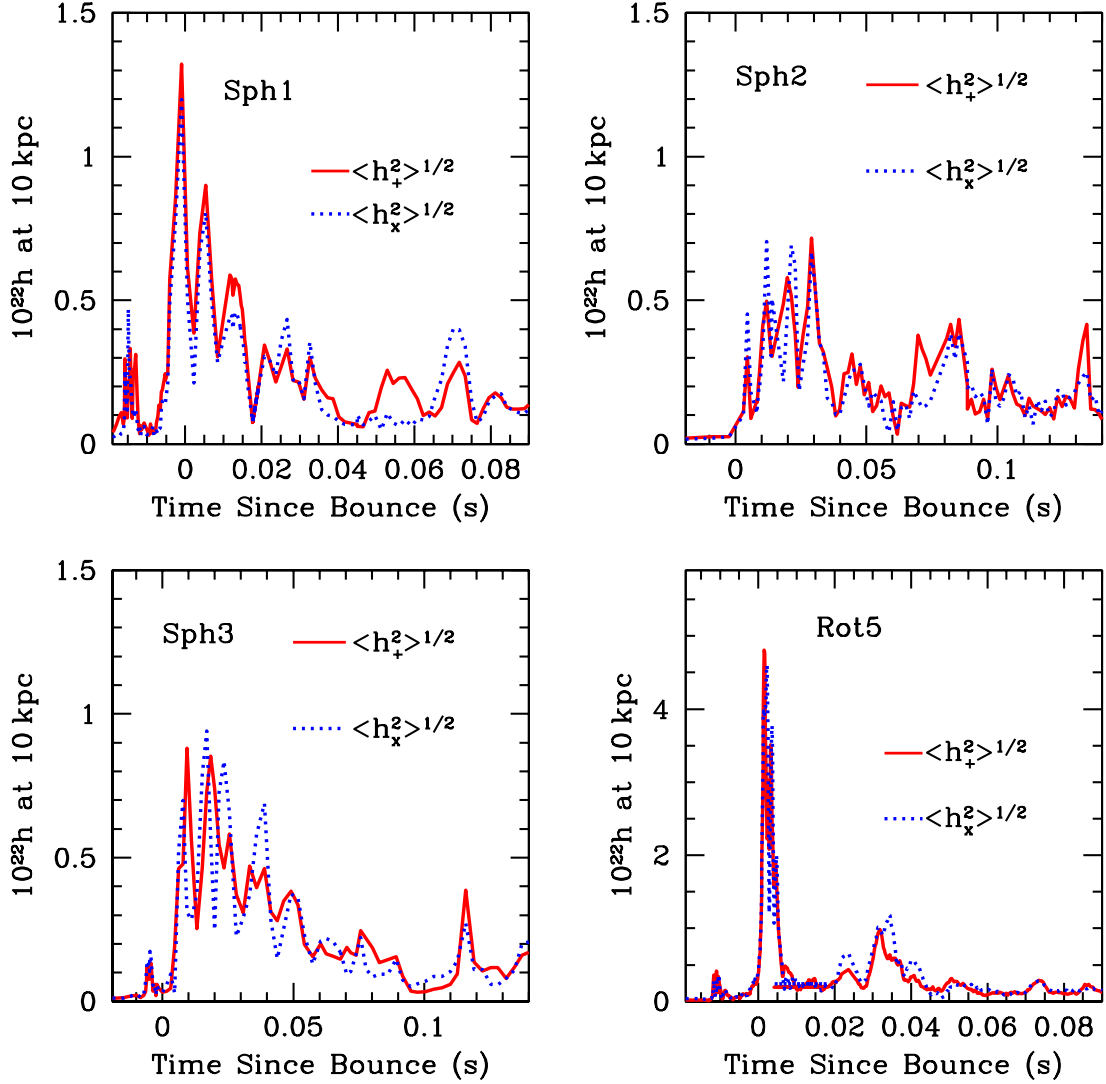}
\caption{The angle averaged wave amplitudes ($<h^2_{+}>^{1/2}$: solid
  line,$<h^2_{x}>^{1/2}$: dotted line) for the mass motions from 4
  non-rotating supernova models.  Sph1-Sph3 are 3 different models
  with 3 levels of resolution (0.3,1.0,3.0 million particles).  In
  these simulations, gravity was assumed to be spherically symmetric.
  Rot5 is a rotating progenitor (corresponding to Rot1: see Fig. 1)
  where the velocity angular was set to zero just before collapse, but
  with the collapse followed under full gravity.  The difference in
  signals arises both from full versus spherically symmetric gravity
  and from the different progenitors.}
\end{figure}
\clearpage

\begin{figure}
\plotone{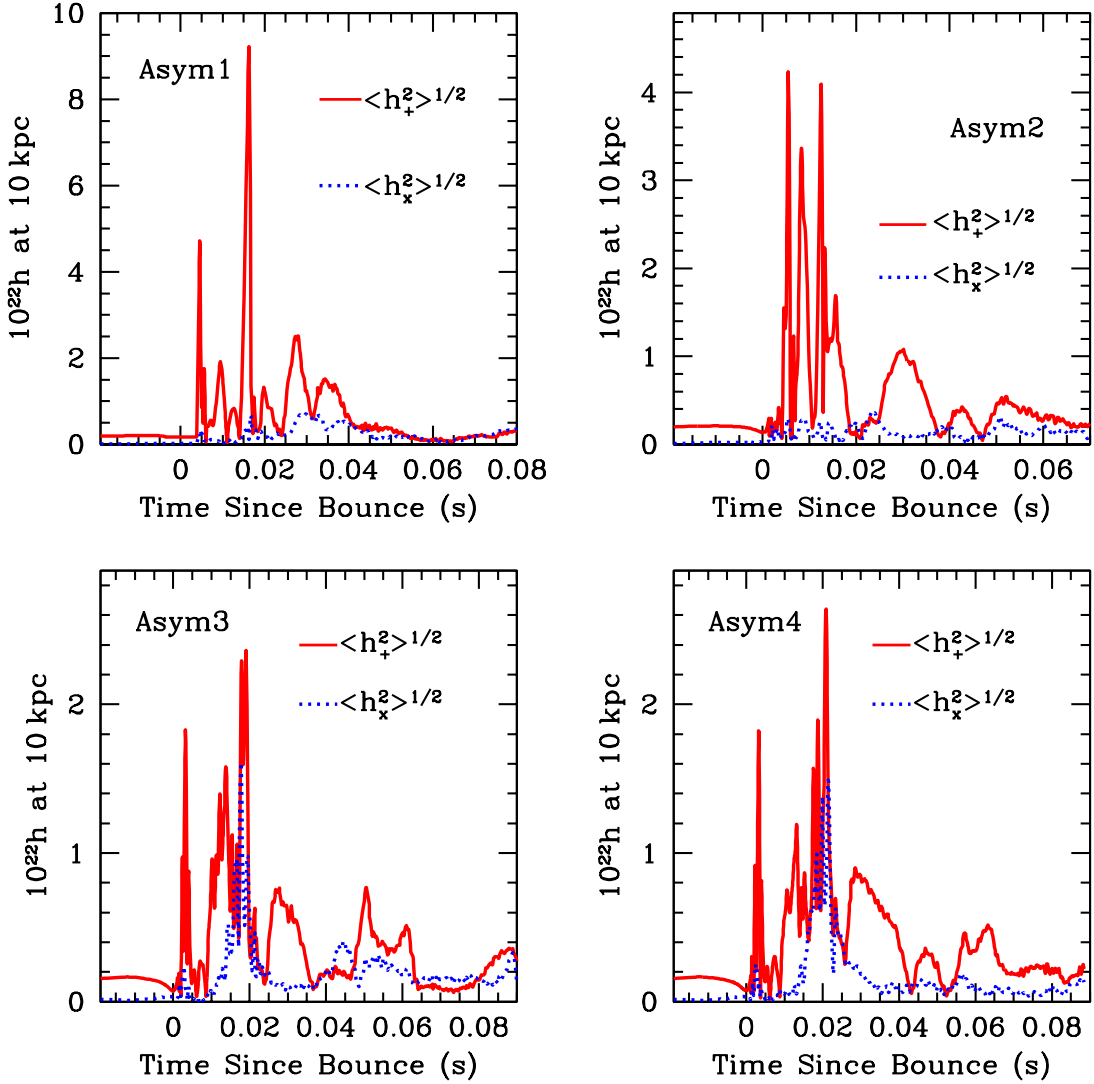}
\caption{The angle averaged wave amplitudes ($<h^2_{+}>^{1/2}$: solid
  line,$<h^2_{x}>^{1/2}$: dotted line) for the mass motions from 4
  supernovae with global perturbations prior to collapse.  Asym1
  corresponds to a 25\% global perturbation throughout the entire core
  (assuming oscillatory modes in the neutron star).  Asym2 corresponds
  to a 40\% global perturbation in the burning shells only.  Asym3 and
  Asym4 correspond to 30\% pertubations in the burning shells, sith 
  and without momentum being carried away by neutrinos.}
\end{figure}
\clearpage

\begin{figure}
\plotone{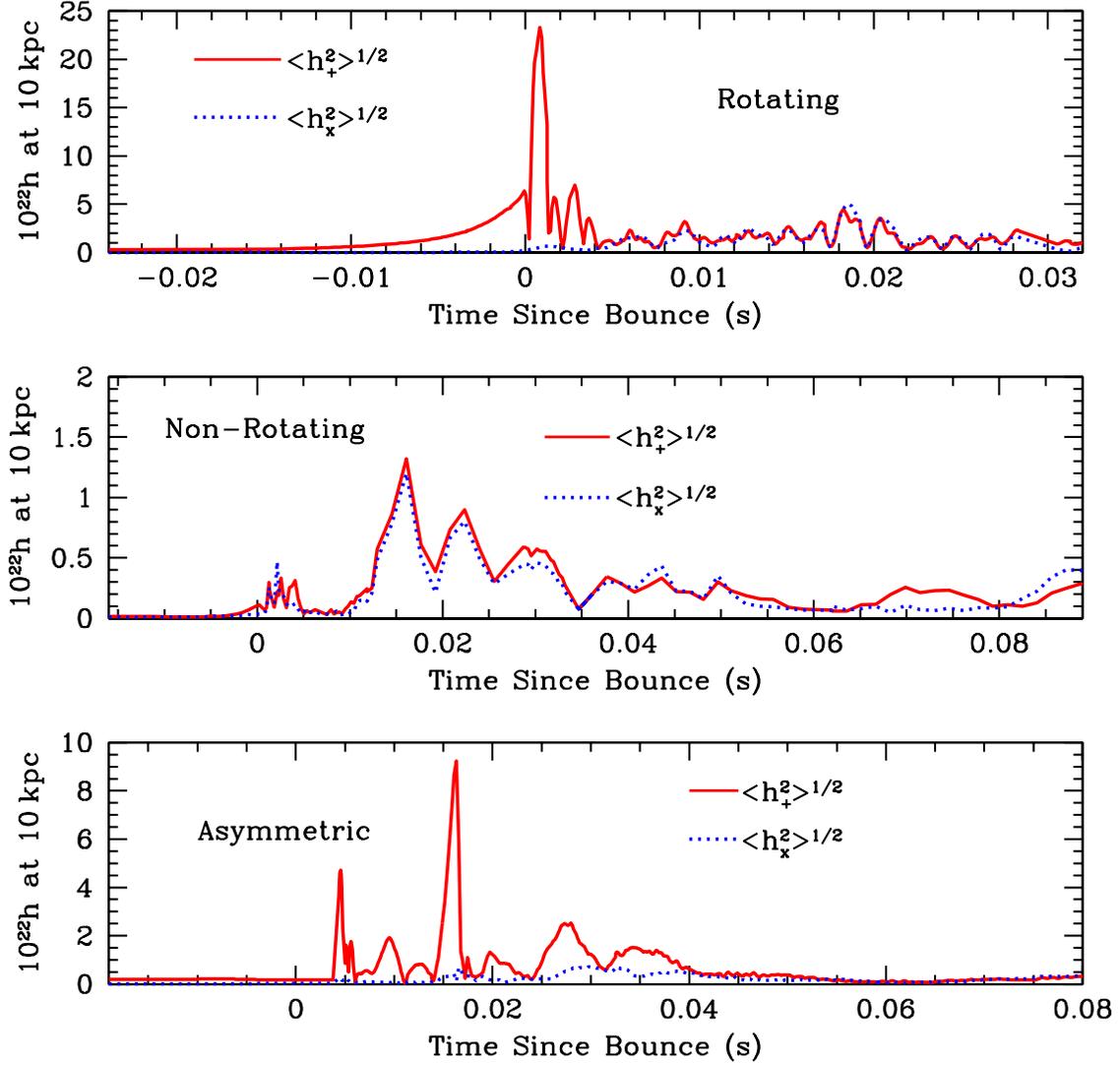}
\caption{The angle averaged wave amplitudes ($<h^2_{+}>^{1/2}$: solid
  line,$<h^2_{x}>^{1/2}$: dotted line) for the mass motions from 3
  representative models of rotating, non-rotating, and asymmetric
  collapse.  The fast-rotator produces the strongest signal which occurs 
  at bounce.  The asymmetric collapse simulation produces a reasonably 
  strong signal, but not necessarily at bounce, and the weak signal from 
  the non-rotating case also does not peak at bounce.}
\end{figure}
\clearpage

\begin{figure}
\plotone{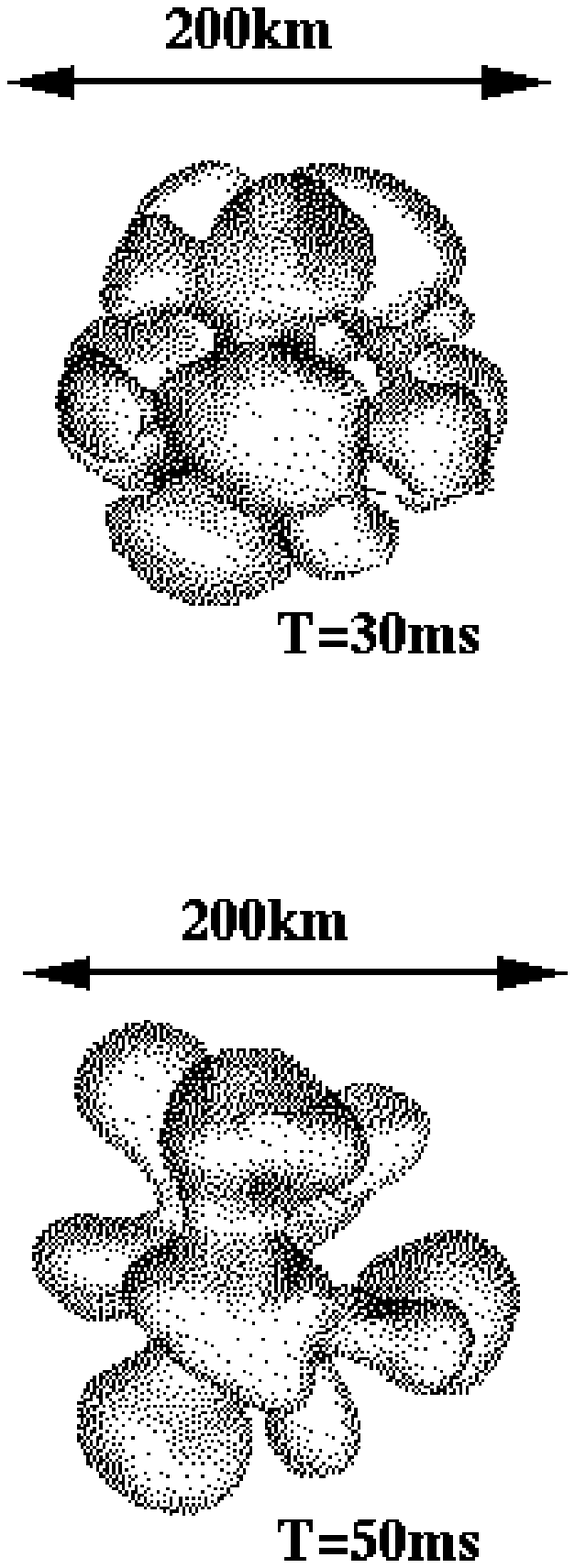}
\caption{Two snapshots in time of the convective upflows in model Sph1.  
As time progresses, the convective modes merge, producing fewer upflows.  
With a sufficiently delayed explosion, it is possible that the modes 
will merge into a single upflow and single downflow, ideal conditions 
to produce a GW signal.}
\end{figure}
\clearpage

\begin{figure}
\plotone{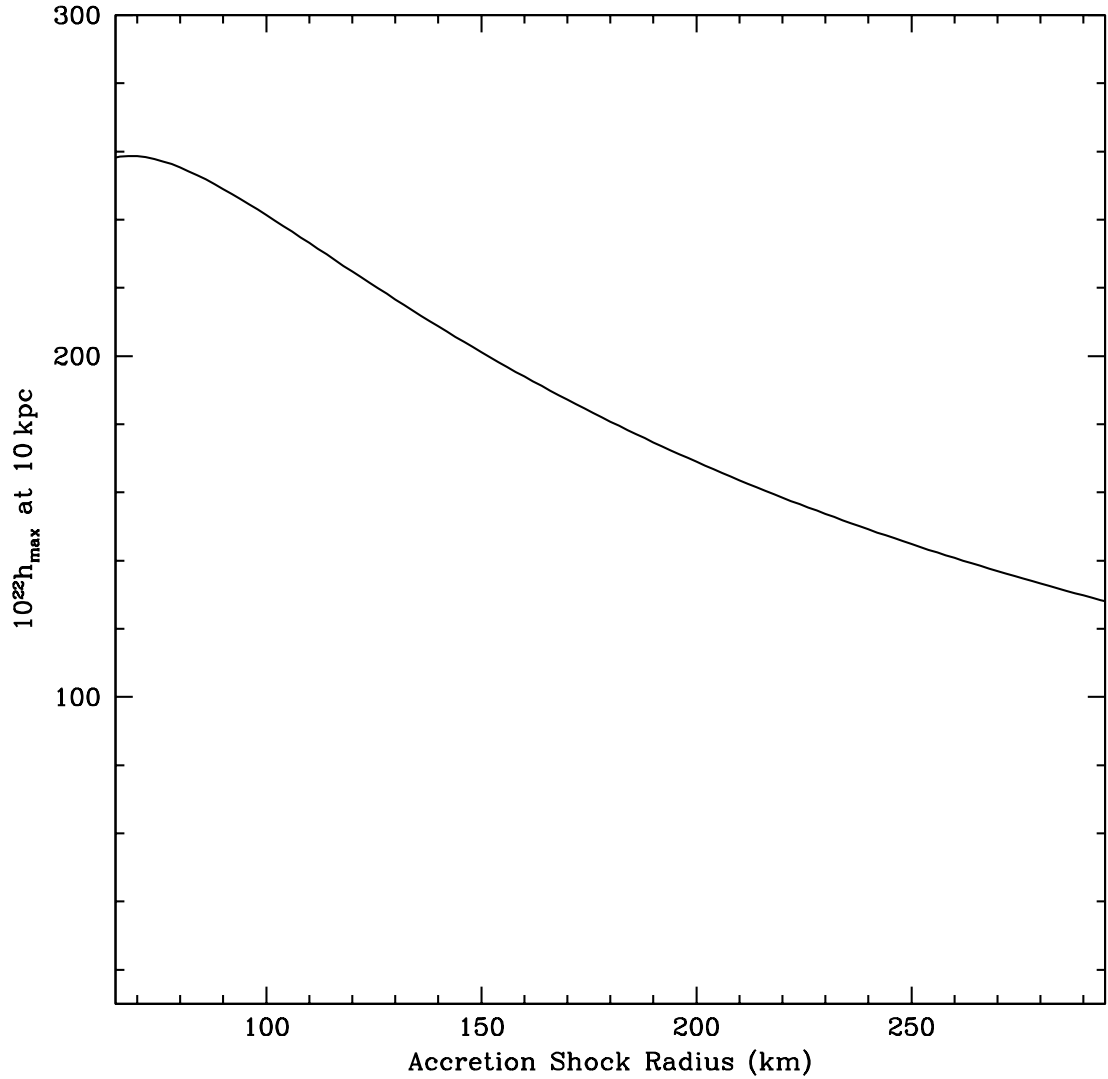}
\caption{Maximum gravitational wave signal from convection as a 
function of the radial extent of the convective region.  As the 
convective region expands, the maximum signal decreases.  Note 
that this estimate is an upper limit, and calculations predict 
signals that are 1--2 orders of magnitude weaker than this value.}
\end{figure}
\clearpage

\begin{figure}
\plotone{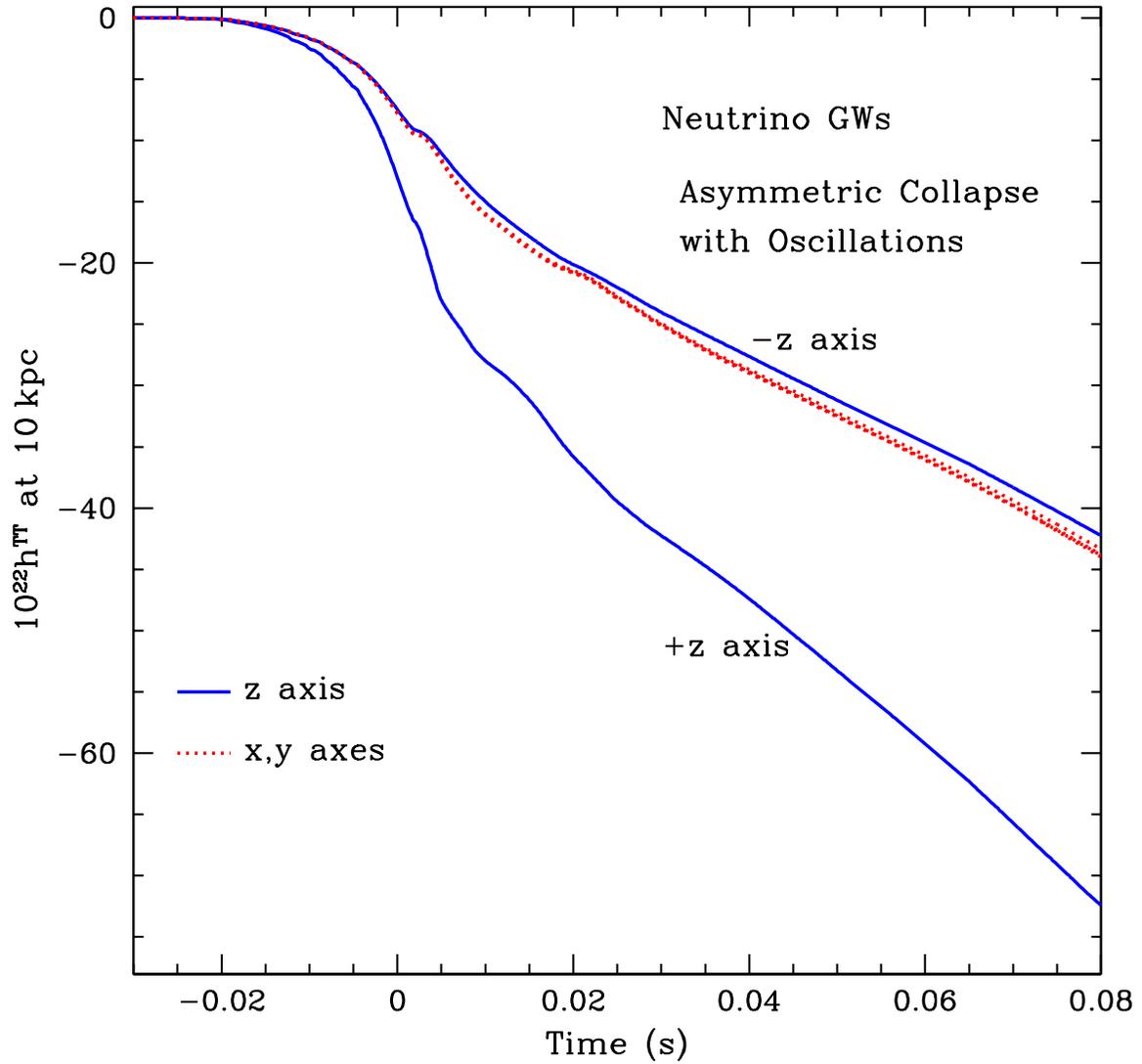}
\caption{Gravitational wave signal for the simulation with a 25\% core 
oscillation  pertubation as a function of time and observer location.  
The signals observed off the perturbation axis are nearly all identical 
and are bracketed by the positive and negative $z$ axis observations.}
\end{figure}
\clearpage

\begin{figure}
\plotone{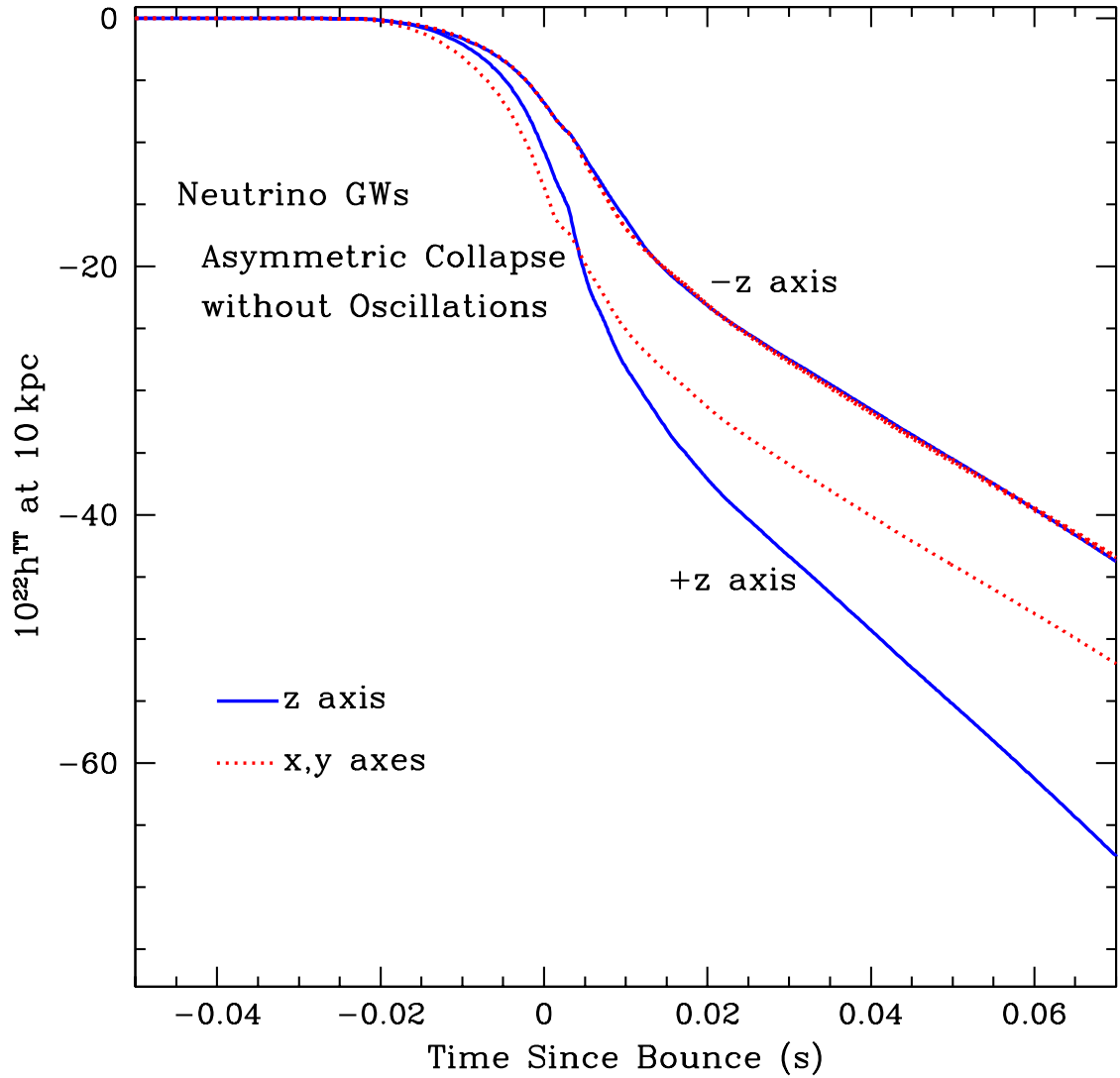}
\caption{Gravitational wave signal for the simulation with a 40\% 
  burning shell pertubation as a function of time and observer
  location.  As with Fig. 7, the signals observed off the perturbation
  axis are nearly bracketed by the positive and negative z axis
  observations.}
\end{figure}
\clearpage

\begin{figure}
\plotone{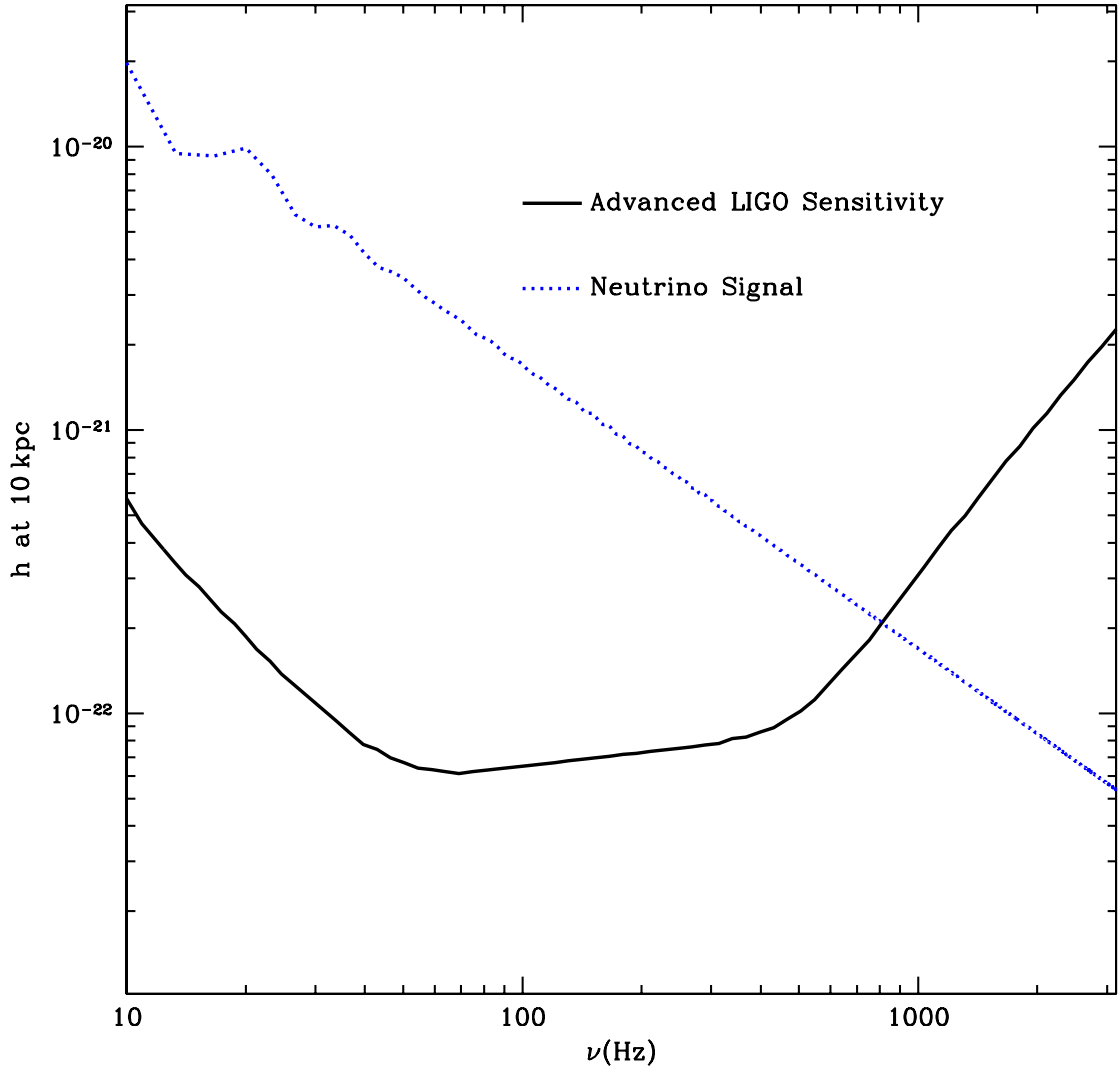}
\caption{Spectral energy distribution (calculated by taking the Fourier 
  transform of the data from fig. 7) of the GW radiation from the
  neutrino asymmetries for model Asym1 (25\% perturbations throughout
  the star).  The advanced LIGO noise curve is plotted for comparison
  (Gustafson et al. 1999).  The neutrino signal for more reasonable
  asymmetries is likely to be an order of magnitude lower.}
\end{figure}
\clearpage

\begin{figure}
\plotone{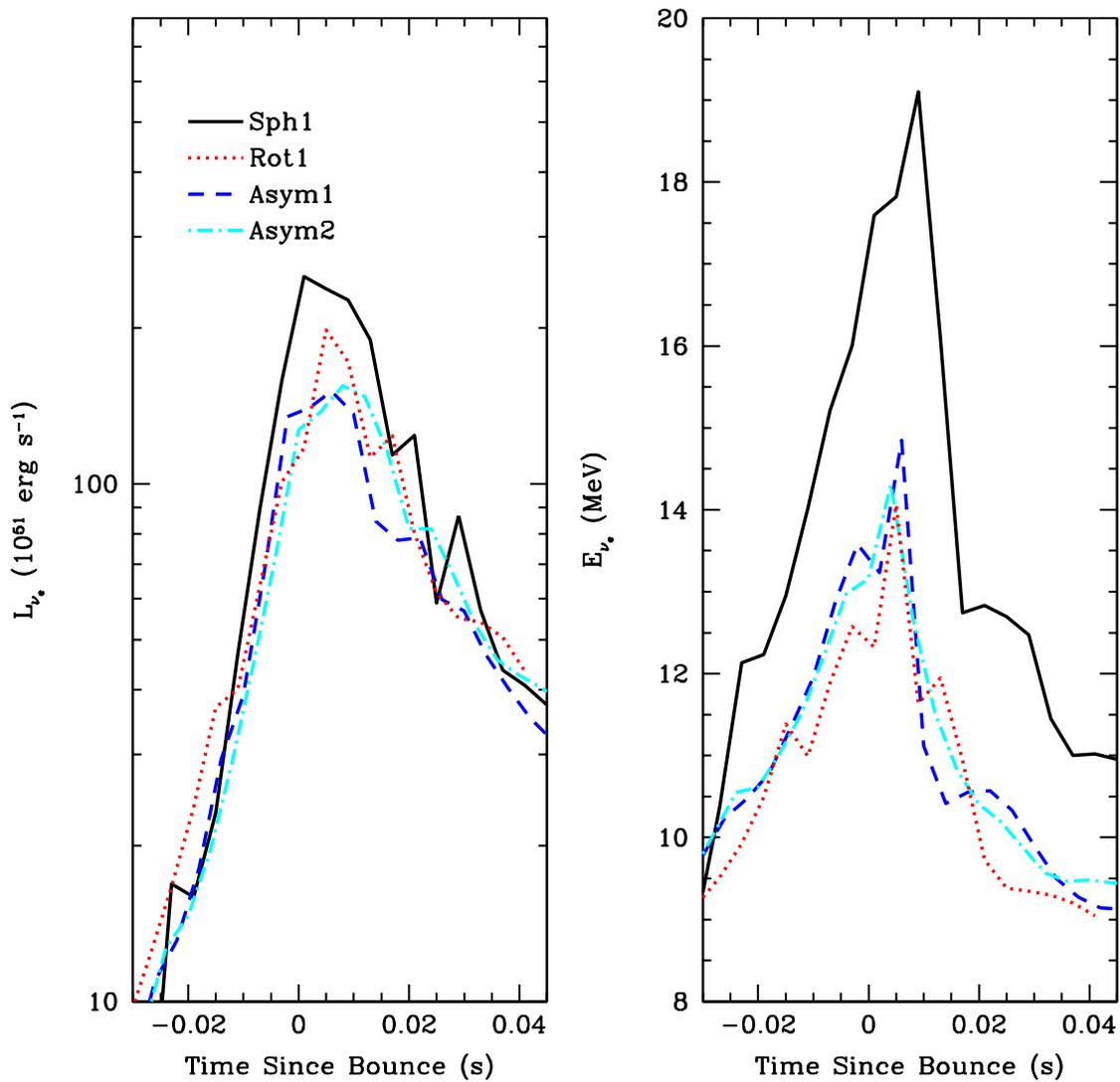}
\caption{Electron neutrino luminosity and energy as a function of 
time since bounce. The luminosities (and energies to the 
30\% level) are all very similar.  Based upon these simulations, it would 
be difficult to distinguish the mass motions from observations of 
just the the electron neutrinos.}
\end{figure}
\clearpage

\begin{figure}
\plotone{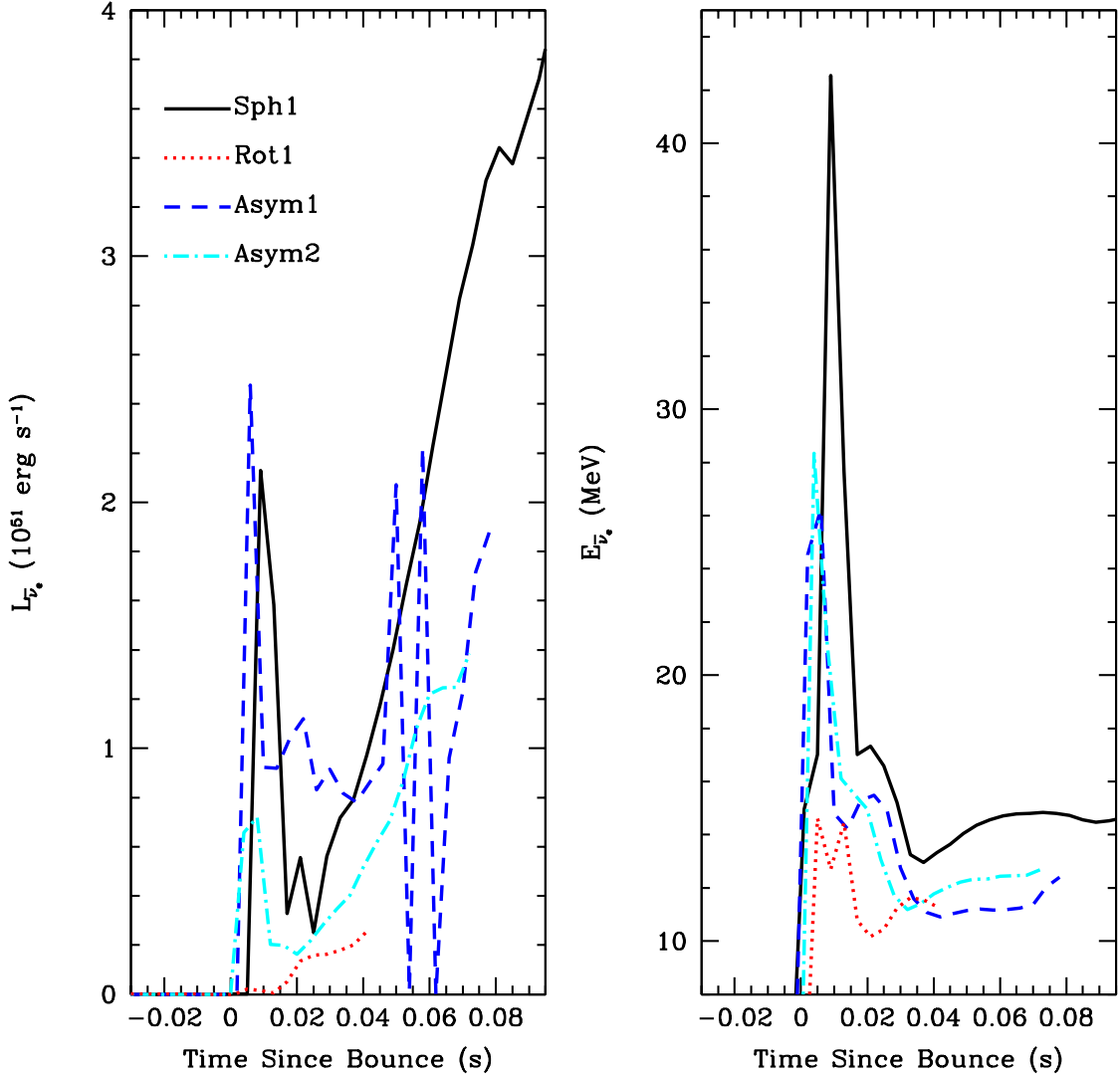}
\caption{Anti-electron neutrino luminosity and energy as a function of 
  time since bounce.  The differences in the luminosities and energies
  are greater for these neutrinos than those of the electron
  neutrinos (Fig. 10), but the uncertainties (due to the lower fluxes) are
  higher for these models.  Neutrino observations may be able to
  constrain the mass motions with more detailed 3-dimensional neutrino
  estimates.}
\end{figure}
\clearpage

\begin{figure}
\plotone{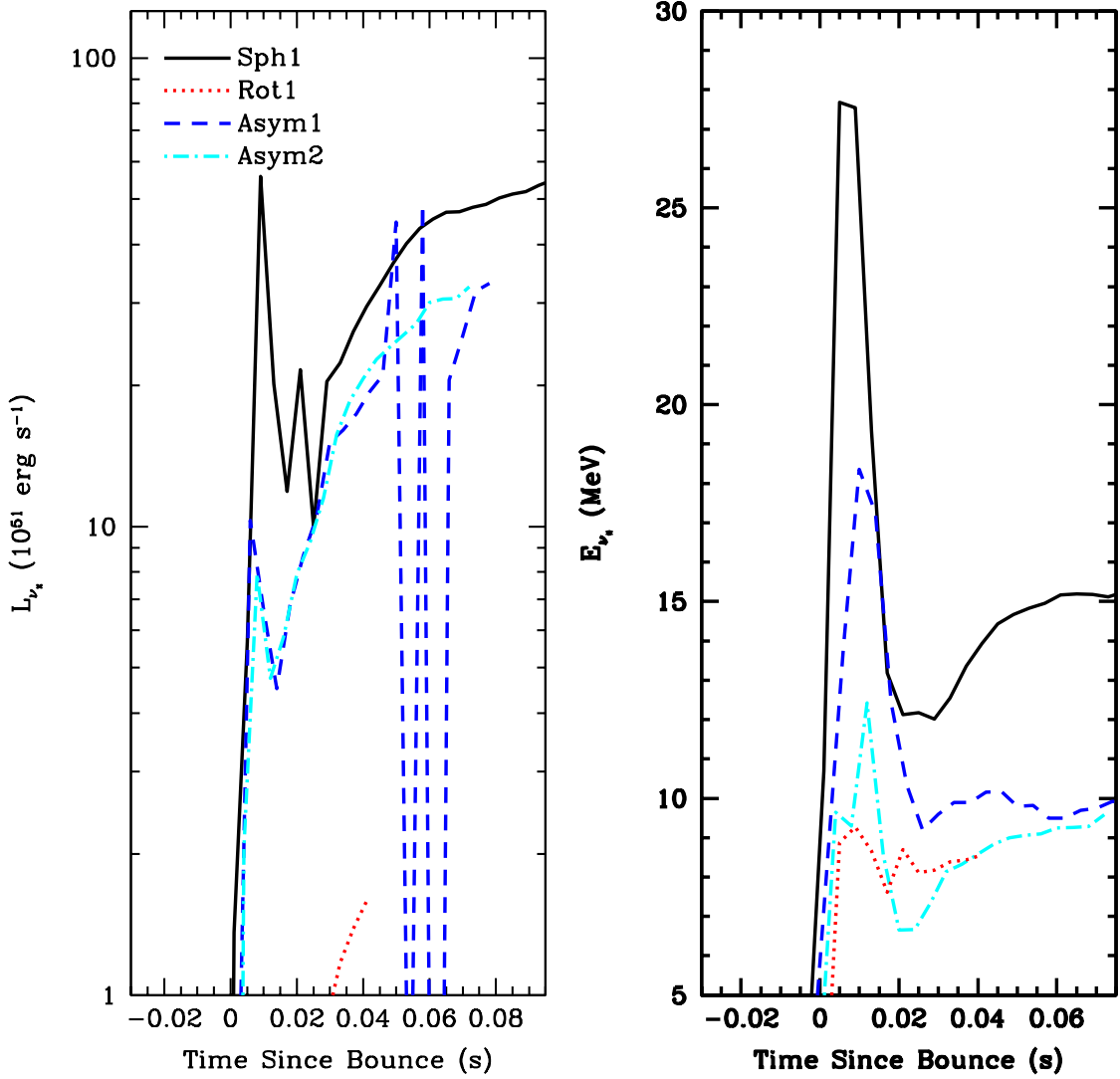}
\caption{$\mu, \tau$ neutrino luminosity and energy as a function of 
  time since bounce.  As with the anti-electron neutrinos, the
  differences in the luminosities and energies are greater for these
  neutrinos than those of the electron neutrinos (Fig. 10), but the
  uncertainties (due to the lower fluxes) are higher for these models.
  Anti-electron neutrino observations are easier to observe, and their
  observations will probably place stronger constraints on the mass 
motions and equation of state. etailed 3-dimensional neutrino estimates.}
\end{figure}
\clearpage

\begin{figure}
\epsscale{.80}
\plotone{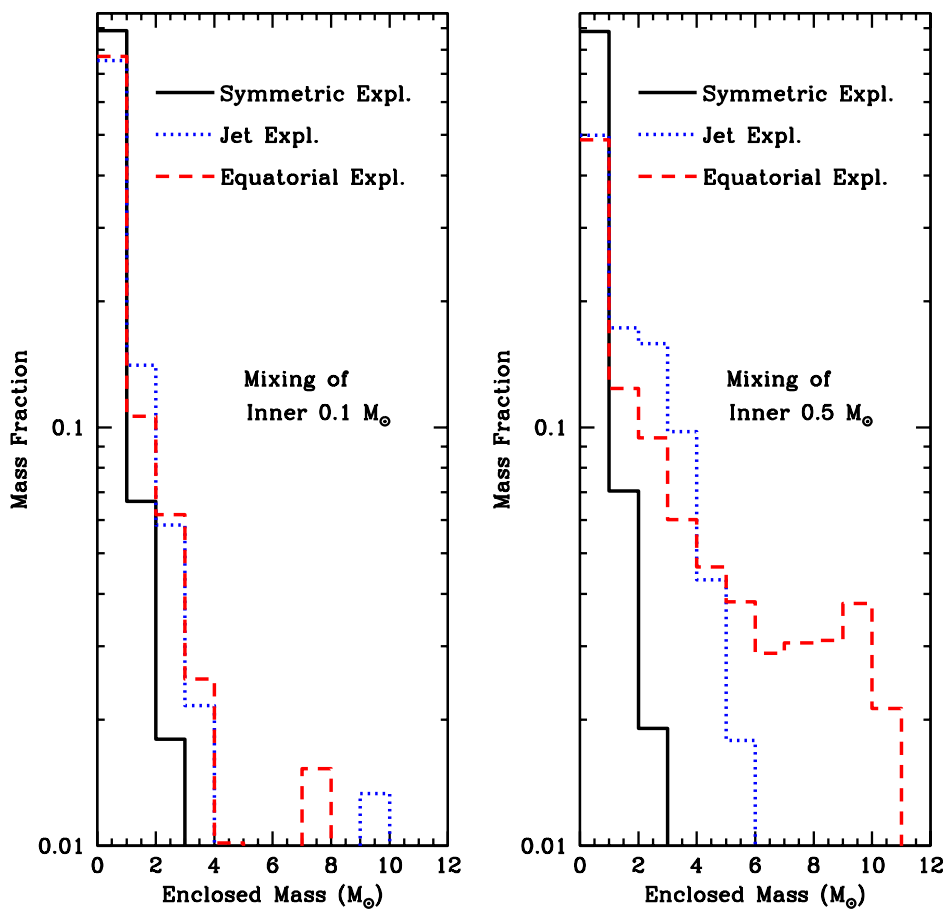}
\caption{Amount of mixing of the inner 0.1\,M$_\odot$ (left) 
  and the inner 0.5\,M$_\odot$ (right) of supernova ejecta for a
  symmetric explosion with decay energy added in (solid line), a polar
  explosion with a jet 2 times stronger along the poles than along the
  equator (dotted lines), and an equatorial explosion with the
  explosion 4 times stronger in the equator than along the poles
  (dashed line).  See Hungerford et al. (2003) for details.}
\end{figure}
\clearpage

\begin{figure}
\epsscale{1}
\plotone{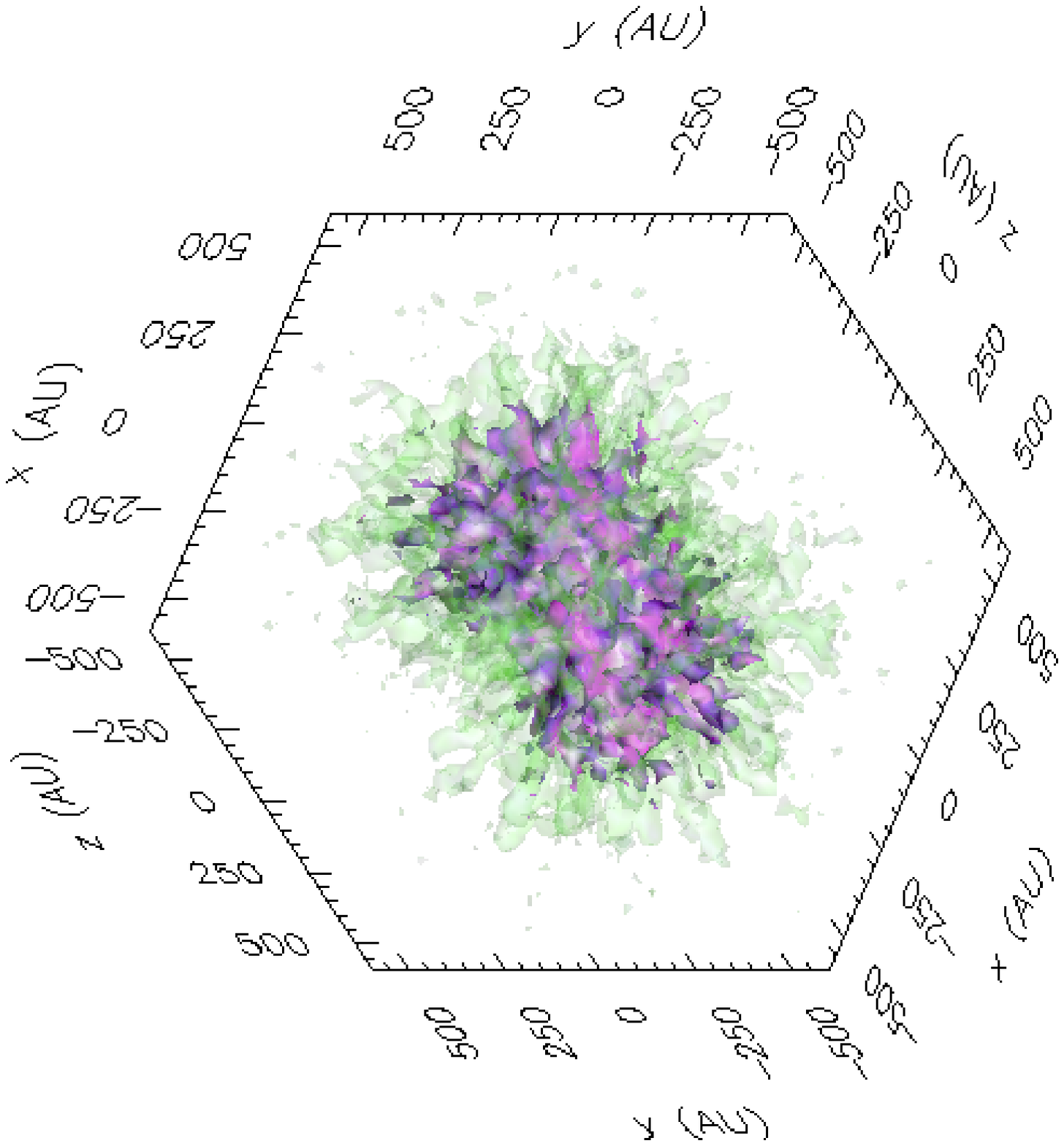}
\caption{Two contours of the nickel distribution for an asymmetric 
  explosion.  The inner (outer) contours correspond to a nickel
  densities of $10^5$ ($10^{-16}$) particles per cm$^{-3}$.  This
  asymmetry mixes the nickel much further out (Fig. 13), altering the
  nucleosynthetic yields.  It also changes the gamma-ray signal
  dramatically and the level and direction of asymmetry can be easily
  distinguished for a Galactic supernova (Hungerford et al. 2003).}
\end{figure}
\clearpage

\end{document}